\pdfoutput=1
\documentclass[reprint,aps,pre,amsmath,amssymb,superscriptaddress,nofootinbib,floatfix]{revtex4-2}

\usepackage{graphicx}
\usepackage[T1]{fontenc}
\usepackage{hyperref}
\hypersetup{colorlinks=true,linkcolor=blue,citecolor=blue,urlcolor=blue}

\begin{document}

\title{Bright Compton source based on the interaction of an ultraintense laser pulse with a relativistically transparent foil and a plasma mirror}

\author{D.~A.~Chekmareva}
\affiliation{A.~V.~Gaponov-Grekhov Institute of Applied Physics of the Russian Academy of Sciences, 46 Ul'yanov Street, 603000 Nizhny Novgorod, Russia}

\author{A.~V.~Korzhimanov}
\email{artem.korzhimanov@ipfran.ru}
\affiliation{A.~V.~Gaponov-Grekhov Institute of Applied Physics of the Russian Academy of Sciences, 46 Ul'yanov Street, 603000 Nizhny Novgorod, Russia}
\affiliation{Faculty of Radiophysics, Lobachevsky State University of Nizhny Novgorod, 23 Gagarin Avenue, 603022 Nizhny Novgorod, Russia}

\begin{abstract}
We propose and numerically study an x-ray source in which a single laser pulse first accelerates
electrons while passing through a relativistically transparent thin overdense foil and then, after
reflection from a plasma mirror, collides head-on with them. Two-dimensional kinetic simulations show
that for a laser pulse with an intensity of the order of $10^{21}$~W/cm$^2$ and optimal foil
parameters, electrons are accelerated to energies of the order of 100~MeV, and the source can deliver
up to $10^{7}$ photons per 0.1\% bandwidth per joule of laser energy in the 10--250~keV range, two
orders of magnitude more than a similar gas-jet scheme. This also exceeds the spectral yield per joule of the
best gas-target betatron sources by about an order of magnitude and is comparable to the best
experimental results obtained with near-critical-density targets.
\end{abstract}

\maketitle

\section{Introduction}

Bright sources of short x-ray pulses in the range from tens of keV to a few MeV are in demand for
phase-contrast imaging of biological objects, radiography, and nuclear resonance fluorescence
spectroscopy~\cite{albert_PPCF_2016}. Laser-plasma electron
acceleration~\cite{tajima_PRL_1979,esarey_RMP_2009} underlies two families of compact femtosecond
sources: betatron radiation of electrons trapped in a wakefield~\cite{kiselev_PRL_2004,rousse_PRL_2004}
and inverse Compton scattering (ICS) of a laser pulse off a counterpropagating beam of relativistic
electrons~\cite{corde_RMP_2013,albert_POP_2023}. In the latter case, the frequency of the scattered
radiation is upshifted by a factor of $\sim4\gamma^2$ (where $\gamma$ is the electron Lorentz factor),
which brings the photons into the range of hundreds of keV and above for electron energies of tens to
hundreds of MeV.

A common drawback of both families remains their relatively low spectral yield. For betatron sources,
it is typically $10^{5}$--$10^{6}$ photons per 0.1\% bandwidth per
shot~\cite{corde_RMP_2013,albert_PPCF_2016}, and only in recent years has it been raised to
$10^{6}$--$10^{7}$ by optimizing the injection and the acceleration
length~\cite{kettle_PRL_2019,wood_PRL_2026,guo_arXiv_2026}. For Compton schemes, the yield is largely
determined by the electron source. In most experimental implementations, this was a gas jet in which
the first pulse drives a wakefield, while a second counterpropagating
pulse~\cite{schwoerer_PRL_2006,chen_PRL_2013} or the same pulse reflected by a plasma mirror behind the
jet~\cite{taphuoc_NP_2012,dopp_PP_2015,yu_SR_2016} scatters off the accelerated electrons. In this way,
photons with energies from a few keV~\cite{schwoerer_PRL_2006} to MeV~\cite{chen_PRL_2013,yu_SR_2016}
and, recently, up to GeV~\cite{matheron_arXiv_2024} have been obtained. Improvements of the plasma
mirror continue to increase the yield~\cite{hu_NP_2026}, but the relatively low electron density in the
jet ($\sim10^{18}$~cm$^{-3}$) limits the number of accelerated, and hence scattering, particles.

The number of radiating electrons can be increased with near-critical-density targets, in which
electrons are accelerated directly by the laser field~\cite{cikhardt_MARAE_2024,lobok_PRE_2021} (the
critical density $N_c=m_e\omega^2/(4\pi e^2)$, where $e$ and $m_e$ are the electron charge and mass and
$\omega$ is the laser frequency, equals $1.74\times10^{21}$~cm$^{-3}$ at a wavelength
$\lambda=800$~nm), and reflection of the pulse from their boundary is accompanied by generation of
radiation extending to the x-ray range~\cite{mikheytsev_MRE_2023}. Compton schemes based on solid and
near-critical targets have been studied theoretically mainly at much higher fields, corresponding to
intensities $I\gtrsim10^{22}$~W/cm$^2$ and normalized amplitudes $a_0\gtrsim50$
($a_0=eE_0/(m_e\omega c)$, where $c$ is the speed of light and $E_0$ is the field amplitude), as sources
of MeV gamma rays~\cite{nerush_PP_2014,vyskocil_PPCF_2020,huang_NJP_2019,liu_PP_2019,jirka_SR_2020,long_PPCF_2019,ji_PPCF_2019,babjak_arXiv_2026}
(a comparative analysis of such configurations is given in Ref.~\cite{gu_MRE_2019}). Recently, a
two-layer ``transparent layer + reflecting substrate'' scheme has also been realized
experimentally~\cite{shou_NP_2022}: at an intensity of $\approx10^{21}$~W/cm$^2$ and a pulse energy of
26~J, a target consisting of a 40-$\mu$m-thick carbon-nanotube layer with an electron density of
$0.4N_c$ and a nanometer-thick foil attached behind it yielded more than $5\times10^{10}$ photons above
10~keV per joule of laser energy at a conversion efficiency of the order of $10^{-3}$. Unlike
Ref.~\cite{jirka_SR_2020}, where a thin solid foil in front of the transparent layer only steepens the
leading edge of the pulse, in the present work the thin foil itself serves as the electron source: its
density is one to two orders of magnitude above critical, and its thickness is tens of nanometers
instead of tens of micrometers, so that electron acceleration and scattering of radiation off the
electrons occur over a length of less than 10~$\mu$m. At the same time, the laser intensity, of the
order of $10^{21}$~W/cm$^2$, corresponds to already operating petawatt facilities and yields photons in
the range of tens to hundreds of keV, which is most in demand for imaging.

\begin{figure}
\centering
\includegraphics[width=\columnwidth]{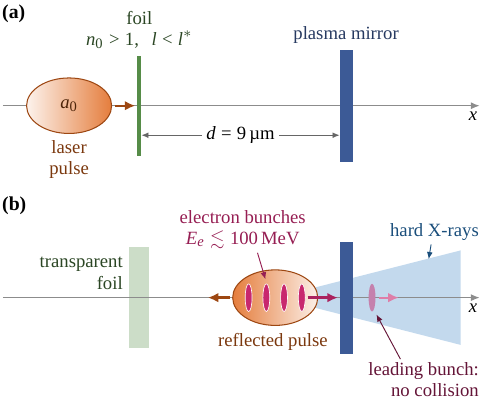}
\caption{Scheme of the source. (a)~Initial configuration: a laser pulse of amplitude $a_0$ is incident
on a foil with the normalized electron density $n_0>1$ and thickness $l$ below the transparency threshold
$l^*$ [Eq.~(\ref{eq:lstar})]. A flat plasma mirror is placed at a distance $d\simeq10~\mu$m behind the
foil. (b)~The transmitted pulse is reflected by the mirror and collides with the accelerated electrons.}
\label{fig:scheme}
\end{figure}

The possibility of using a foil with a density above critical as an electron source while letting the
pulse pass through it is provided by the effect of relativistically induced transparency of a thin
layer~\cite{vshivkov_PP_1998,korzhimanov_Zh_2007,korzhimanov_PvZh_2007,gonoskov_PRL_2009,eremin_PP_2010,miheycev_Ke_2020}:
if the foil thickness is below a threshold, the light pressure pushes out the electron layer before the
electrostatic charge-separation field can hold it, and the pulse passes through the target together
with part of the electrons. Below, we propose an ICS source scheme in which a single pulse is used
twice: first it accelerates electrons while making the foil transparent, and then, after reflection from
a flat plasma mirror behind the foil, it collides head-on with them (Fig.~\ref{fig:scheme}). The scheme
combines a solid-state electron density with no need to synchronize two pulses.

\section{Source scheme and simulation setup}

In a head-on collision of an electron with Lorentz factor $\gamma\gg1$ with a wave of amplitude
$a\gg1$, scattering is nonlinear, of wiggler type, with the parameter $K=a$~\cite{corde_RMP_2013}: the
frequency of the $n$th harmonic, corresponding to the absorption of $n$ laser photons, is
$4\gamma^2n\omega/(1+a^2/2)$, and the radiation is a synchrotron-like continuum with the critical
frequency $\omega_c\approx3a\gamma^2\omega$, i.e., $n_c\approx(3/8)a^3$. For $\lambda=800$~nm
($\hbar\omega=1.55$~eV), a reflected-pulse amplitude $a\approx20$--25, and $\gamma\approx30$--100, this
gives $\hbar\omega_c\approx0.1$--1~MeV and $n_c\sim(3$--$6)\times10^3$, whereas linear Thomson
scattering ($n=1$) yields photons with energies below 1~keV and plays no role. The quantum parameter
$\chi_e\approx2\gamma a\hbar\omega/(m_ec^2)\approx1.2\times10^{-3}a$ for $\gamma=200$. Hence, for
existing laser systems the emission is classical, and pair production is negligible.

Radiation reaction effects, which are determined by the accumulated energy loss of the electron during
the collision, are likewise small. The fraction of energy lost by an electron passing through a pulse
of duration $\tau$ and envelope amplitude $a$ is
$\Delta\mathcal{E}/\mathcal{E}\approx(2\pi r_e/\lambda)\,\omega\tau\,\gamma a^2\approx2\times10^{-8}\,\omega\tau\,\gamma a^2$
($r_e$ is the classical electron radius), so that for intense femtosecond pulses ($\omega\tau\approx60$
and $a\approx20$) and electrons with $\gamma$ of the order of 100 it amounts to about 5\%. The ratio of the
radiation reaction force to the Lorentz force does not exceed
$(4/3)(2\pi r_e/\lambda)\gamma^2a\approx0.03$.

Let the foil have an electron density $N_e=n_0N_c$ and a thickness $l$. At normal incidence of the
pulse on a layer with a sharp boundary, the electrons are pushed by the light pressure, while the ions
remain nearly immobile during the pulse (the effect of ion mobility on the transparency threshold is
discussed in Ref.~\cite{miheycev_Ke_2020}). The layer becomes transparent when the period-averaged light
pressure exceeds the maximum force with which the electrostatic charge-separation field can hold the
electrons. For a circularly polarized wave of amplitude $a_0$, this condition reads~\cite{vshivkov_PP_1998}
$a_0>\varepsilon_0$, where $\varepsilon_0=\pi n_0 l/\lambda$ is the dimensionless areal density of the
layer. Since the light pressure is determined by the intensity, for a linearly polarized wave, whose
amplitude $a_0$ is $\sqrt2$ times larger at the same intensity, the transparency condition becomes
$a_0/\sqrt2>\varepsilon_0$, i.e.,
\begin{equation}
l<l^*=\frac{a_0}{\sqrt2\,\pi n_0}\,\lambda .
\label{eq:lstar}
\end{equation}
For $l<l^*$, the electrostatic field $E_x\simeq2\pi eN_el$ can no longer hold the displaced electron
layer, and the pulse passes through the foil together with part of the
electrons~\cite{korzhimanov_Zh_2007}. The parameter $\varepsilon_0$ also determines the linear
reflection coefficient of the layer, $\varepsilon_0^2/(1+\varepsilon_0^2)$~\cite{vshivkov_PP_1998}.
Therefore, for $\varepsilon_0\lesssim1$ the layer is transparent already for a weak wave, and
criterion~(\ref{eq:lstar}) is meaningful only for $\varepsilon_0>1$, when transparency requires
$a_0>\sqrt2\,\varepsilon_0>1$. This is precisely the region of special interest, because the number of
electrons in the layer can be large there, but the foil can then become transparent only at a
relativistic amplitude, when the light pressure considerably changes the thickness of the electron
layer. This mechanism, called relativistically induced slab transparency in
Ref.~\cite{gonoskov_PRL_2009}, differs from relativistic self-induced transparency of an extended plasma,
which is due to the relativistic increase of the electron inertia and occurs for a linearly polarized
wave at $n_0\lesssim\sqrt{1+a_0^2/2}$~\cite{siminos_PRE_2012,siminos_NJP_2017}. We stress that
criterion~(\ref{eq:lstar}) refers to the instantaneous field amplitude: for a pulse with a smooth
envelope, a layer of thickness $l$ becomes transparent at the moment when $a(t)$ reaches
$\sqrt2\,\pi n_0 l/\lambda$, so that the leading part of the pulse is reflected and the rest is
transmitted~\cite{wang_PRL_2011}.

The simulations were performed with the PICADOR code~\cite{surmin_CPC_2016} in two-dimensional
geometry. Photon emission was described by a Monte Carlo module~\cite{gonoskov_PRE_2015}, in which each
macroparticle emitted with the local synchrotron probability depending on $\chi_e$, and the momentum of
the emitted photon was subtracted from the electron momentum. Thus, both radiation reaction and the
potentially quantum character of the emission were taken into account in the simulations, despite the
smallness of these effects. The pulse had a Gaussian envelope in time and space, a duration of 11~fs,
and was focused into a spot 5~$\mu$m in diameter (both quantities are full widths at half maximum of the
intensity) in the foil plane. It was linearly polarized, with the polarization plane at $45^\circ$ to
the simulation plane, and had an energy of 5~J. Thus, its peak amplitude was $a_0=26.7$, corresponding
to an intensity of $1.5\times10^{21}$~W/cm$^2$ and a power of 430~TW. Such parameters are attainable at
typical modern petawatt-class laser systems~\cite{korzhimanov_Ufn_2011} with additional pulse
post-compression~\cite{ginzburg_OE_2021,sladkov_P_2023}.

The foil was modeled as a fully ionized layer with a sharp boundary. Its thickness $l$ and
normalized density $n_0$ were varied. The ions were mobile with a charge-to-mass ratio of
$3e/7m_p$ ($m_p$ is the proton mass), which corresponds to helium-like iron ions
$^{56}$Fe$^{24+}$ (this ionization state corresponds to field ionization at an intensity of
$\approx10^{21}$~W/cm$^2$~\cite{korzhimanov_PRL_2012,korzhimanov_Ke_2013}). The mirror was a plasma
layer with an electron density of $57N_c$ and a thickness of 0.8~$\mu$m located at a distance
$d=9~\mu$m behind the foil.

The simulation box was $25\times20~\mu$m$^2$ with a grid step $\Delta x=12.5$~nm along the propagation
axis and $\Delta y=100$~nm across it, with 100 particles per cell. Absorbing boundary conditions were
used along $x$ and periodic ones along $y$. The scan over $(n_0,l)$ covered $n_0=10$--90 and
$l=(0.025$--$0.4)\lambda$.

\section{Electron acceleration from a foil undergoing relativistic transparency}

\begin{figure}
\centering
\includegraphics[width=\columnwidth]{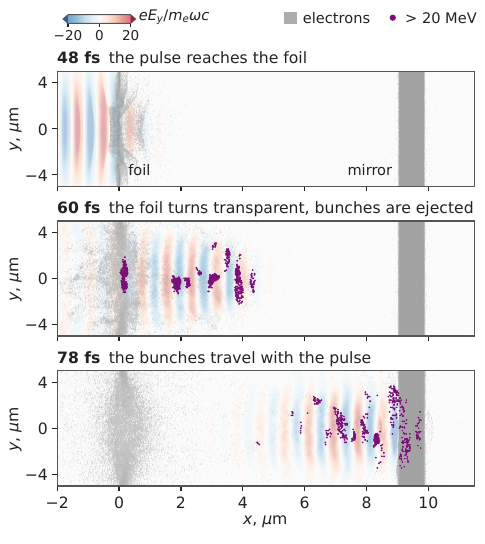}
\caption{Interaction of the pulse with the foil for $n_0=55$, $l=0.06\lambda$ at $t=48$, 60, and 78~fs
(from top to bottom): transverse field $E_y$ (color scale), electrons (gray), and electrons with
energies above 20~MeV (magenta dots).}
\label{fig:interaction}
\end{figure}

Figure~\ref{fig:interaction} shows the typical dynamics of the laser pulse interaction with a thin
foil. Upon reaching the foil, the pulse pushes the electrons toward its rear boundary by the light
pressure, so that the thickness of the electron layer becomes smaller than the skin depth, and the laser
pulse passes through it. This is accompanied by simultaneous dragging of part of the electrons out of
the plasma layer~\cite{eremin_PP_2010}. The ions do not have time to move. The electrons leave the layer
as a train of bunches separated by $\lambda/2$, which moves together with the pulse toward the mirror.
Some electrons are also accelerated backward, but to considerably lower energies.

The electrons that have left the layer are accelerated directly by the field of the transmitted wave,
and their energy is determined by the transverse momentum $p_\perp$ gained in it:
$\gamma\approx(1+p_\perp^2/m_e^2c^2)/2R$, where $R=\gamma-p_x/m_ec$ is conserved during motion in a
plane wave ($R=1$ for an electron initially at rest). Since $p_\perp\sim am_ec$, where $a$ is the field
amplitude at the electron position, energies of the order of 100~MeV are attainable only for electrons
that find themselves in the region of the field maximum. The role of the foil is precisely to inject
electrons into this region: it holds them until the most intense part of the pulse arrives. Without the
foil, the electrons are swept away by the leading edge of the pulse and never reach the field maximum.
The analysis of electron trajectories (Supplemental Material, Fig.~\ref{fig:tracks}) confirms this
picture: all electrons with energies above 45~MeV are torn out of the foil when the pulse maximum passes
through it and gain their energy from the transverse wave field, whereas in a simulation without the
foil the energy of test electrons does not exceed 25~MeV.

Direct acceleration of electrons from an ultrathin foil by a tightly focused pulse was considered
earlier in Ref.~\cite{vais_QE_2017}. Experimentally, post-acceleration of electrons ejected from a thin
polymer foil, driven by the radiation transmitted through the foil, was observed in Ref.~\cite{braenzel_PRL_2017} at an
intensity of $\sim10^{20}$~W/cm$^2$, where a second foil at a distance of 500~$\mu$m merely cut off the
pulse, preventing electron deceleration in the decaying field. In our scheme, the second target serves
as a mirror, and the pulse reflected by it is used for scattering, while the distance of
$\sim10~\mu$m allows the electrons to meet it before they slip behind the transmitted pulse.

\begin{figure*}
\centering
\includegraphics[width=0.86\textwidth]{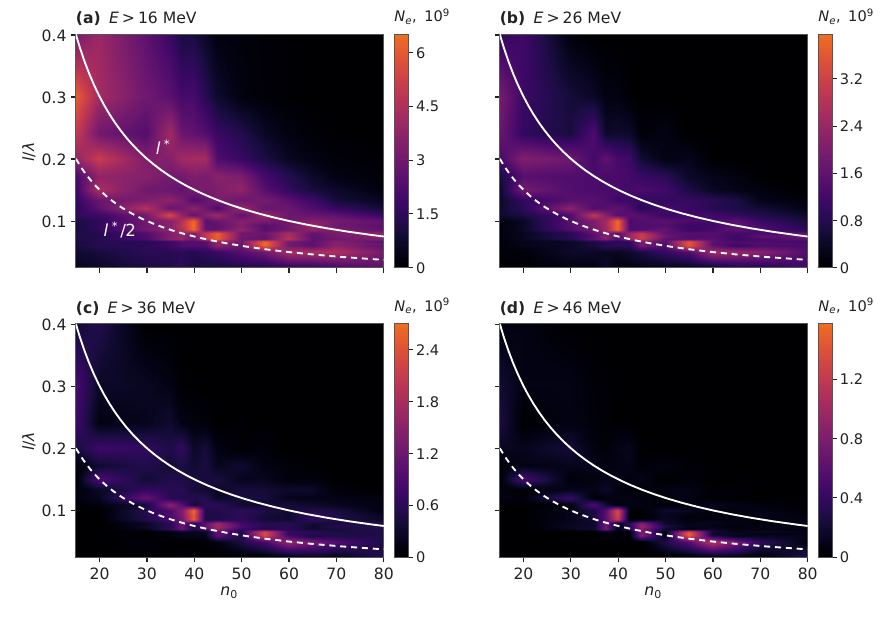}
\caption{Number of electrons $N_e$ with energies above (a)~16, (b)~26, (c)~36, and (d)~46~MeV moving
in the pulse propagation direction in the plane of foil parameters $(n_0,l)$ for $a_0=26.7$ at
$t=72$~fs (linear interpolation between simulations, the size along the third coordinate is taken to be
5~$\mu$m). Solid line: transparency threshold $l^*(n_0)$ [Eq.~(\ref{eq:lstar})], dashed line: $l^*/2$.}
\label{fig:maps}
\end{figure*}

To find the optimal target parameters, we constructed maps of the number of electrons above a given
energy in the $(n_0,l)$ plane at a fixed time $t\approx72$~fs, when the acceleration is almost complete
and the particles have not yet left the simulation box (Fig.~\ref{fig:maps}). For thresholds above
$\sim25$~MeV, the high-energy population is concentrated in a narrow band that follows the dependence
$l\propto1/n_0$ and lies between $l^*/2$ and $l^*$: for $n_0\geq30$, the maximum over thickness is
reached on average at $l\approx0.6\,l^*$. The largest number of electrons above 26~MeV,
$\approx4\times10^9$, was obtained at $n_0=40$, $l=0.1\lambda$, and above 46~MeV,
$\approx1.6\times10^9$, at $n_0=55$, $l=0.063\lambda$. Note that within this band the dependence on
the foil parameters is not smooth: in simulations with close values of $n_0$ and $l$, the number of
high-energy electrons can differ by a factor of 2--3. Electrons with energies above 16~MeV are observed
on both sides of the transparency boundary: this is a quasi-thermal component produced by heating at
the target surface, with a temperature of the order of the ponderomotive energy
$m_ec^2(\sqrt{1+a_0^2/2}-1)\approx9.2$~MeV. Above the threshold, $l>l^*$, the pulse is reflected from
the foil, and the accelerated electrons form a diffuse cloud without a collimated train of bunches.

The position of the optimum can be explained by the dependence of the transparency criterion on the
instantaneous amplitude noted above. For $l\ll l^*$, the layer becomes transparent already at the weak
leading edge of the pulse and stops holding the electrons long before the field maximum arrives, and,
moreover, the number of electrons in it is small. For $l\to l^*$, transparency sets in only near the
pulse maximum, most of the pulse energy is reflected, and the transmitted pulse is short and weak, so
that both the number of electrons torn out of the layer and their energy are small. The optimum is
reached at an intermediate thickness $l\approx l^*/2$: on the one hand, transparency sets in at the
pulse front at $a(t)\approx a_0/2$, and a substantial part of the pulse passes through the layer, and on
the other hand, the layer keeps holding some of the electrons until the field maximum arrives.

The spectra of the accelerated electrons are nonthermal but broad and show no formation of
quasi-monoenergetic beams. This is natural for acceleration by the wave field: the electrons are
injected into the wave at different phases, and their energy changes as they slip behind the wave. The
spectral cutoff energy reaches its maximum, $\approx95$~MeV, at the same parameters as the number of
high-energy electrons, $l\approx0.5\,l^*$ (Supplemental Material, Fig.~\ref{fig:espectra}). As the
optimal parameters, we chose $n_0=45$--55 and $l\approx l^*/2$, which for $\lambda=800$~nm corresponds
to $N_e\approx(0.8$--$1)\times10^{23}$~cm$^{-3}$ and thicknesses of 44--53~nm. The results below refer
to $n_0=55$, $l=0.06\lambda$ ($l\approx0.55\,l^*$), unless stated otherwise.

\section{X-ray generation}

When the laser pulse reaches the mirror, it is reflected and starts propagating toward the accelerated
electrons. However, some of the electrons located at the pulse front manage to fly further without
taking part in the collision with the reflected radiation. As a result, the simulations show that only
about 80\% of the electrons with energies above 25~MeV are involved in the ICS process. This is
confirmed by the photon distributions (Supplemental Material, Fig.~\ref{fig:collision}): radiation with
energies above 100~keV is not observed for the electrons at the leading edge and appears only for the
electrons following behind.

\begin{figure*}
\centering
\includegraphics[width=\textwidth]{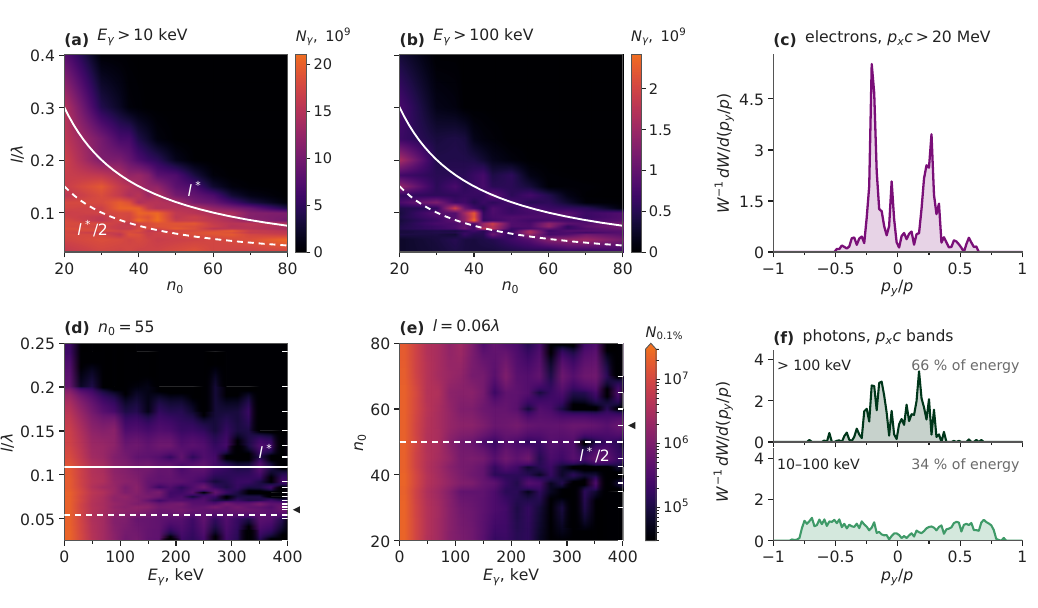}
\caption{Forward-moving photons produced in the collision of the electrons with the pulse reflected
from the mirror (increment between $t=78$ and 96~fs). (a),~(b)~Number of photons $N_\gamma$ with
energies above 10 and 100~keV in the plane of foil parameters $(n_0,l)$. (d),~(e)~Photon spectra per
0.1\% bandwidth versus foil thickness at $n_0=55$ and versus $n_0$ at $l=0.06\lambda$. Tick marks on the
right indicate the simulations, and triangles mark $l=0.06\lambda$ and $n_0=55$. Solid lines:
$l=l^*$ [Eq.~(\ref{eq:lstar})], dashed lines: $l=l^*/2$. (c),~(f)~Unit-normalized energy distributions over
$p_y/p=\sin\vartheta$ in the simulation plane at $n_0=55$, $l=0.06\lambda$: (c)~electrons with
$p_xc>20$~MeV ($t=78$~fs), (f)~photons in two $p_xc$ intervals, with the fraction of the photon energy
in each interval indicated on the right. In panels (a), (b), (d), and (e), linear interpolation between
simulations is used, the size along the third coordinate is taken to be 5~$\mu$m, and the number of
photons is estimated from the distribution of the radiated energy.}
\label{fig:phspec}
\end{figure*}

The analysis of the photon distributions in time and space shows that photons moving in the pulse
propagation direction are generated in three stages in the scheme under consideration. At the first
stage ($t\approx48$--66~fs), radiation arises from the interaction of the pulse with the foil and is
directed mainly backward. The second stage ($t\approx78$--96~fs) is the actual collision of the accelerated
electrons with the pulse reflected from the mirror, and the photons are generated near the mirror
surface (Fig.~\ref{fig:collision} of the Supplemental Material). Finally, the third stage is observed at
$t\approx108$--126~fs, when the reflected pulse returns to the foil: photons are generated near its
initial position and also move forward. Figure~\ref{fig:phspec} shows the characteristics of the
second-stage emission.

The number of second-stage photons above 10~keV (Fig.~\ref{fig:phspec}a) is nearly independent of the
foil parameters in the entire transparency region $l<l^*$ and amounts to $(1.6$--$1.9)\times10^{10}$ per
shot, while for $l>l^*$ it falls off rapidly. Apparently, the soft part of the spectrum is formed by
numerous electrons of relatively low energy, which are present at any parameters of the transparent
foil. The number of photons above 100~keV (Fig.~\ref{fig:phspec}b), in contrast, follows the
distribution of high-energy electrons (Fig.~\ref{fig:maps}): it is maximal in the band
$l\approx(0.5$--$0.7)\,l^*$, where it reaches $(1.5$--$2.4)\times10^{9}$, while outside the band it is
$(0.4$--$0.9)\times10^{9}$.

Like the electron spectrum, the second-stage photon spectrum (Figs.~\ref{fig:phspec}d and
\ref{fig:phspec}e) is broad and decreases monotonically with photon energy: the spectral yield decreases
from $1.5\times10^7$ photons/0.1\% bandwidth at 10~keV to $2\times10^6$ at 100~keV and $3\times10^5$ at
250~keV. The hard part of the spectrum is sensitive to the foil parameters in the same way as the
electron cutoff energy (Supplemental Material, Fig.~\ref{fig:espectra}): for $n_0=55$, the yield of
photons with energies around 300~keV is maximal at $l=(0.06$--$0.067)\lambda$ and several times smaller
at other thicknesses, and for $l=0.06\lambda$ it is maximal at $n_0=45$, 55, and 70.

The total yield of the source is the sum of the emission from all three stages. To convert the results
of the two-dimensional simulations into the total number of photons, the effective size along the third
coordinate was taken equal to the laser beam diameter, 5~$\mu$m. Then the number of photons emitted per shot in the pulse propagation direction ranges from
$\approx5\times10^7$ photons/0.1\% bandwidth at 10~keV to $\approx2\times10^6$ at 250~keV. Per joule of laser energy, this corresponds to a spectral yield of up to
$10^7$ photons/(0.1\% bandwidth$\cdot$J). Note that the emission region of hard photons is narrower than
the laser spot: its effective transverse size decreases from $\approx5~\mu$m for photons around 10~keV
to $\approx2.5~\mu$m for photons around 100~keV. Therefore, for the hard part of the spectrum, the above
values are upper estimates and may be overestimated by about a factor of two. The total number of
forward-moving photons in the 10--250~keV range was $\approx7\times10^{10}$ per shot, and their total
energy was $\approx0.4$~mJ, i.e., the conversion efficiency of laser energy into x rays is
$\eta\approx10^{-4}$. Note that the broad emission spectrum is primarily due to the broad electron
spectrum, since $\omega_c\propto\gamma^2$. It was shown earlier for gas-jet schemes that interaction with
quasi-monoenergetic electron beams also narrows the photon
spectrum~\cite{rykovanov_JPB_2014,khrennikov_PRL_2015,yu_SR_2016}. Obtaining quasi-monoenergetic
electron beams in the scheme considered here, however, remains an open problem.

Another feature of the source is the strong dependence of the angular distribution of the radiation on
the photon energy (Figs.~\ref{fig:phspec}c and \ref{fig:phspec}f). Hard photons with $p_xc>100$~keV,
which carry two thirds of the energy emitted at the second stage, follow the two-lobe distribution of the
electrons over $p_y/p$, and 80\% of their energy is emitted at $|p_y/p|<0.25$, whereas softer photons
are distributed almost uniformly up to $|p_y/p|\approx0.7$ ($\vartheta\approx45^\circ$). This is a
consequence of the wiggler regime: the radiation is emitted at angles within $K/\gamma$ in the
polarization plane~\cite{corde_RMP_2013}. Therefore, the hardest photons, emitted by the most energetic
electrons, are concentrated within a small angle, while the soft ones are spread over a much larger
angle (see the Supplemental Material for details).

The source size for the second-stage emission is set by the transverse size of the radiating part of
the electrons at the moment of collision, $\approx3~\mu$m, and the flash duration by its length, i.e.,
$\tau\approx5$--10~fs, since the photons are emitted forward and move together with the electrons. When
estimating the solid angle $\Delta\Omega$ into which the radiation is emitted, one must take into
account that it is not axially symmetric: the electrons oscillate in the polarization plane, and the
photons are emitted into a fan elongated along it, while the distributions in Fig.~\ref{fig:phspec}f
represent its projection onto the simulation plane. An estimate of the length and thickness of the fan
(Supplemental Material) gives the solid angle containing 80\% of the radiated energy ranging from
$\approx1.2$~sr for photons with energies of 10--100~keV to $\approx0.13$~sr for photons above 100~keV.
Hence, the peak brightness $B=N_{0.1\%}/(\tau A\,\Delta\Omega)$ estimated with $\tau=10$~fs,
$A=(3~\mu\mathrm{m})^2$, and the above second-stage spectral yield is $\approx10^{20}$
photons/(s$\cdot$mm$^2\cdot$mrad$^2\cdot$0.1\%) at photon energies of 15--100~keV and
$\approx2\times10^{19}$ at 250~keV. Thus, the brightness is comparable to that of gas-jet plasma-mirror
sources at similar photon energies~\cite{taphuoc_NP_2012} and about an order of magnitude lower than that
of the quasi-monochromatic MeV source of Ref.~\cite{yu_SR_2016}.

\section{Discussion and conclusions}

The main advantage of the proposed scheme is the high spectral yield of photons per joule of laser energy
in the 10--250~keV range. For the structurally identical gas-jet scheme with a plasma mirror, it is
$(2$--$4)\times10^{4}$ photons per 0.1\% bandwidth per joule at 50--200~keV~\cite{taphuoc_NP_2012},
i.e., one to two orders of magnitude lower. Betatron sources deliver $10^{5}$--$10^{6}$ photons per
0.1\% bandwidth per shot~\cite{corde_RMP_2013,albert_PPCF_2016}, or $10^{4}$--$10^{5}$ per joule for
pulse energies of 1--15~J, and the best recent experiments deliver $4\times10^{5}$ at
5~keV~\cite{kettle_PRL_2019} and $\approx2\times10^{6}$ at 10--20~keV (our estimate based on the on-axis
flux of $7\times10^{4}$ photons/(mrad$^2\cdot$0.1\%), the divergence of $17\times16$~mrad, and the pulse
energy of 10.5~J reported in Ref.~\cite{wood_PRL_2026}). The proposed scheme outperforms them by about an
order of magnitude at 15~keV and by much more at 100~keV, where the spectrum of betatron radiation with a
critical energy of $\approx20$~keV falls off exponentially.

Closest to the proposed scheme in yield are sources based on near-critical-density targets. Betatron
radiation in a foam at a pulse energy of 60--80~J contains $\sim3\times10^{11}$ photons above
5~keV~\cite{cikhardt_MARAE_2024}, but its spectrum is much softer and the conversion efficiency is
$\sim10^{-5}$. Simulations of the relativistic self-trapping regime of a 135-TW pulse yielded
$3\times10^{10}$ photons above 10~keV~\cite{lobok_PRE_2021}, which is close to our value. In the
experiment of Ref.~\cite{shou_NP_2022}, the spectral intensity was
$\approx4\times10^{6}$ photons/(sr$\cdot$0.1\%$\cdot$J) at 15 and 100~keV, whereas in our simulations,
taking into account the solid angle of the emission from all three stages ($\approx1$~sr above 10~keV
and $\approx0.4$~sr above 100~keV), it is $\approx10^{7}$ and $\approx5\times10^{6}$, respectively.
Thus, the spectral yield per joule of the scheme is at the level of the best results for near-critical
targets with a fivefold lower pulse energy and an interaction length below 10~$\mu$m instead of 40--80~$\mu$m,
although in conversion efficiency ($10^{-4}$ versus $10^{-3}$) it is inferior to the scheme of
Ref.~\cite{shou_NP_2022}, whose spectrum extends to MeV energies.

The gain in spectral yield, however, does not translate into peak brightness: the source emits into a
solid angle about three orders of magnitude larger than, e.g., the $\sim1^\circ$ cone in
Ref.~\cite{taphuoc_NP_2012}. The comparison is therefore most meaningful for applications that collect a
large solid angle, such as contact and projection radiography and phase-contrast imaging of
centimeter-sized objects at a short distance from the source. Per shot of a 5-J laser, the source
yields $\approx7\times10^{10}$ photons with energies of 10--250~keV into a solid angle of $\approx1$~sr,
i.e., $\sim10^4$ photons per $100\times100~\mu$m$^2$ detector pixel at a distance of 30~cm, which is
sufficient for single-shot radiography. The compact Compton source MuCLS based on a storage ring
provides $\approx10^{10}$ photons/s at 15--35~keV within a 3--4\% bandwidth with a source size of
$\approx40~\mu$m~\cite{eggl_JSR_2016}, whereas the proposed source delivers
$\sim(2$--$3)\times10^{10}$ photons in the same range per $\sim10$-fs flash with a source size of a few
micrometers. Therefore, at a laser repetition rate of $\sim1$~Hz, the total photon flux is comparable to
that of MuCLS, although the spectral flux density is about an order of magnitude lower. The main
niche of the source is single-shot radiography and phase-contrast imaging of fast processes. We also
note that these estimates were obtained in two-dimensional geometry, and the absolute numbers depend on
the assumed effective extent along the third axis, which, however, is fixed by the pulse energy. The
comparison of the simulation results with experimental data is therefore only an estimate.

Both limitations, the broad spectrum and the large divergence, are properties of the electron source
rather than of the scattering geometry. Narrowing the spectrum requires control of the electron
injection phase, e.g., by structuring the rear surface of the foil or by introducing a longitudinal
density gradient. The divergence is determined by the transverse momentum of the electrons in the wave
field, and reducing it would increase the brightness, $B\propto1/\Delta\Omega$, more efficiently than
increasing the number of photons. Reducing $K$ at the scattering stage (by weakening or defocusing the
reflected pulse) would also narrow the angular distribution of the radiation toward $1/\gamma$ at the
cost of the photon number. These options were not studied in the present work.

From the experimental point of view, the optimal target has a reduced density compared to solids, but
its areal density $N_el\approx4\times10^{17}$~cm$^{-2}$ equals that of a fully ionized carbon foil
$\approx6$~nm thick. Free-standing diamond-like carbon films thinner than 5~nm have already been used in
laser ion acceleration experiments~\cite{ma_NIMPRA_2011}, and it would only be necessary to expand such a
film by about an order of magnitude in thickness while preserving its areal density, which in principle
can be achieved with a controlled prepulse~\cite{macchi_RMP_2013}. An alternative could be
carbon-nanotube foams of the required density~\cite{bin_PRL_2015,wang_H_2021}, which have already been
used in experiments on electron and gamma-ray generation~\cite{rosmej_NJP_2019,shou_NP_2022}.

In summary, we have proposed and numerically studied a single-pulse ICS source in which the gas-jet
electron source is replaced by a solid foil that becomes transparent under the action of the pulse. We
have shown that at $I\approx10^{21}$~W/cm$^2$ the optimal foil parameters are $n_0=45$--55 and
$l\approx l^*/2$, for which the foil injects electrons into the field maximum of the transmitted pulse,
where they are accelerated by the wave field to energies of the order of 100~MeV. The source delivers up
to $10^7$ photons per 0.1\% bandwidth per joule of laser energy around 10~keV and up to $4\times10^5$ at
250~keV. The conversion efficiency of laser energy into the 10--250~keV range is $\approx10^{-4}$, and
the peak brightness reaches $\sim10^{20}$ photons/(s$\cdot$mm$^2\cdot$mrad$^2\cdot$0.1\%). In terms of
the spectral yield per joule, the source exceeds the best gas-target betatron sources by about an order
of magnitude and is comparable to sources based on near-critical-density targets. The scheme requires no
capabilities beyond those already demonstrated at existing petawatt facilities and appears to be a
promising route toward brighter compact sources of hard x rays.

\medskip
\noindent\textit{Supplemental Material.} Electron trajectory analysis, maps of electron spectra, electron and
photon phase portraits at the collision stage, and details of the solid-angle estimate are given in the
Supplemental Material appended after the references.

\begin{acknowledgments}
This work was supported by the Russian Science Foundation (project No.~22-19-00371).
The authors declare no conflicts of interest.
\end{acknowledgments}

\bibliography{compton_en}

\clearpage
\onecolumngrid
\begin{center}
{\large\bfseries Supplemental Material}
\end{center}
\twocolumngrid

\setcounter{figure}{0}
\renewcommand{\thefigure}{S\arabic{figure}}
\setcounter{section}{0}
\renewcommand{\thesection}{S\arabic{section}}

Equations and figures without the prefix S refer to the main text. The notation is the same:
$n_0=N_e/N_c$ is the normalized electron density of the foil, $l$ is its thickness,
$l^*=a_0\lambda/(\sqrt2\,\pi n_0)$ is the transparency threshold~(\ref{eq:lstar}), $a_0=26.7$, and
$\lambda=800$~nm.

\section{Electron trajectory analysis}

\begin{figure}
\centering
\includegraphics[width=\columnwidth]{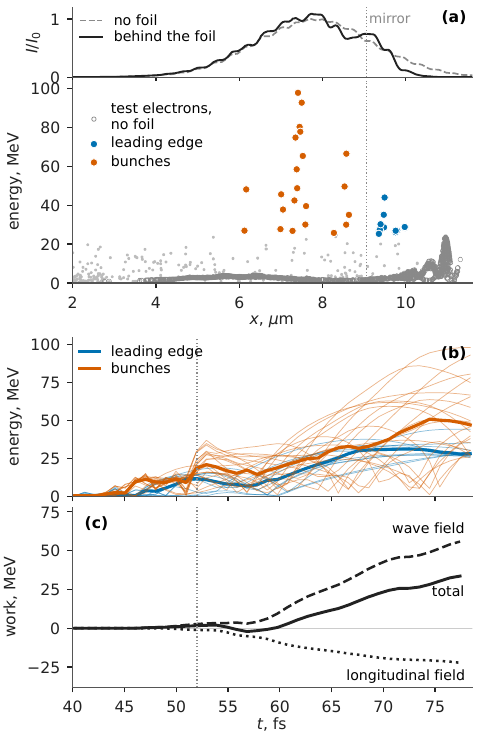}
\caption{Trajectories of foil electrons in simulations with $n_0=45$--55, $l=(0.06$--$0.075)\lambda$.
(a)~Time $t=78$~fs: period-averaged on-axis intensity behind the foil and without it (top) and energy of
the tracked electrons versus coordinate (bottom). Electrons with energies above 25~MeV at the leading
edge of the pulse and in the bunches are highlighted in color, and circles show test electrons in the
simulation without collective fields. (b)~Energy of the highlighted electrons versus time, with thick
lines showing the medians over the groups. (c)~Work of the transverse wave field and of the longitudinal
field after leaving the foil, averaged over the bunch electrons. The vertical dotted line in (b) and (c)
marks the arrival of the pulse maximum at the foil.}
\label{fig:tracks}
\end{figure}

To verify the mechanism of electron injection into the field maximum, electron trajectories were
recorded in simulations with plasma parameters close to the optimal ones ($n_0=45$--55,
$l=(0.06$--$0.075)\lambda$) (Fig.~\ref{fig:tracks}). By the time the pulse reaches the mirror
($t=78$~fs), the electrons with energies above 25~MeV form two groups (Fig.~\ref{fig:tracks}a). The
first group is located at the leading edge of the transmitted pulse: these electrons left the foil at
the moment it became transparent, 6--11~fs before the arrival of the field maximum, and their energy does
not exceed 45~MeV. The second group consists of bunches ejected from the foil when the most intense part
of the pulse passed through it and located near the pulse maximum. All electrons with energies above
45~MeV belong to this group. For comparison, Fig.~\ref{fig:tracks}a shows test electrons from a
simulation in which the electron density of the foil was chosen negligibly small and collective fields
were absent: all of them are gathered at the leading edge of the pulse, and their energy does not exceed
25~MeV.

The energy of the bunch electrons changes nonmonotonically (Fig.~\ref{fig:tracks}b), following the
transverse momentum as the electron slips behind the wave in phase. Splitting the energy gain into the
work of the transverse wave field and the work of the longitudinal field (Fig.~\ref{fig:tracks}c) shows
that after leaving the foil the bunch electrons gain energy only from the wave field (about 50~MeV on
average). The longitudinal charge-separation field decelerates them and takes away about 20~MeV on
average without changing the qualitative picture, while for the leading-edge electrons its work is
negligible. Thus, the role of the foil is to inject electrons into the field maximum, where they are
accelerated directly by the field of the transmitted wave.

\section{Electron spectra}

\begin{figure}
\centering
\includegraphics[width=\columnwidth]{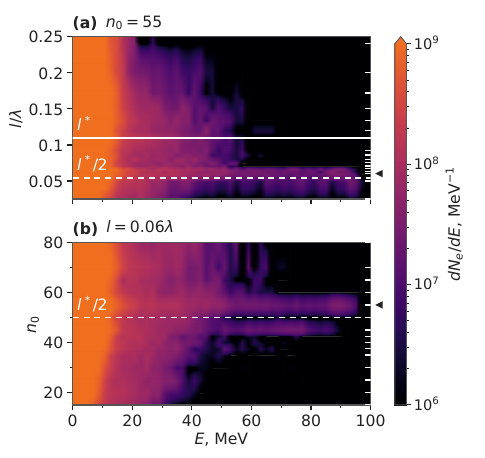}
\caption{Spectra of electrons moving in the pulse propagation direction at $t=75$~fs: (a)~versus foil
thickness at $n_0=55$, (b)~versus $n_0$ at $l=0.06\lambda$ (linear interpolation between simulations,
whose positions are marked by tick marks on the right, the size along the third coordinate is taken to be
5~$\mu$m). Solid line: $l=l^*$, dashed lines: $l=l^*/2$, triangles mark $l=0.06\lambda$ and $n_0=55$.}
\label{fig:espectra}
\end{figure}

The spectra of electrons moving in the pulse propagation direction are shown in
Fig.~\ref{fig:espectra}. They are nonthermal but broad and show no formation of quasi-monoenergetic
beams. The spectra have a pronounced cutoff energy, which reaches its maximum at the same parameters as
the number of high-energy electrons (Fig.~\ref{fig:maps} of the main text). For example, for $n_0=55$
(Fig.~\ref{fig:espectra}a), the spectrum extends to $\approx95$~MeV only in a narrow thickness interval
$l=(0.05$--$0.067)\lambda$, i.e., at $l=(0.46$--$0.61)\,l^*$, while already at $l=0.071\lambda$ the
cutoff energy drops to 55--70~MeV and then slowly decreases with increasing thickness. For
$l=0.06\lambda$ (Fig.~\ref{fig:espectra}b), the cutoff energy grows with $n_0$ from $\approx35$~MeV at
$n_0=15$ to 90--96~MeV at $n_0=45$ and 55, and with a further increase in $n_0$ it decreases again to
65--70~MeV. As in the maps of Fig.~\ref{fig:maps}, this dependence is nonmonotonic: at $n_0=50$ the
cutoff is only $\approx70$~MeV. Thus, the optimum corresponds to the most efficient injection of
electrons into the field maximum. The hard part of the photon spectrum (Figs.~\ref{fig:phspec}d and
\ref{fig:phspec}e of the main text) follows the electron cutoff energy.

\section{Collision stage}

\begin{figure}
\centering
\includegraphics[width=\columnwidth]{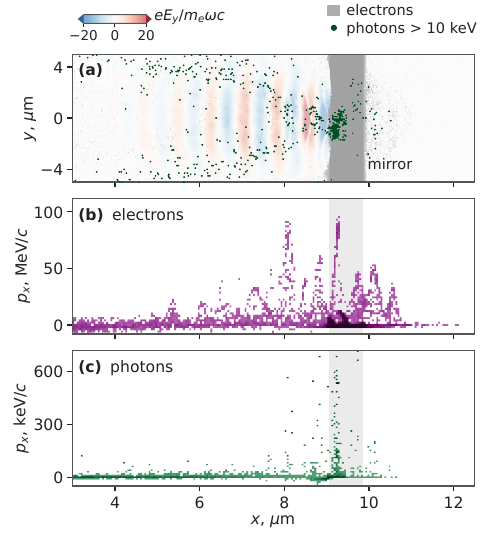}
\caption{Collision stage for $n_0=55$, $l=0.06\lambda$, $t=81$~fs. (a)~Field $E_y$ (color scale),
electrons (gray), and photons with energies above 10~keV (green dots). The mirror is shown by the gray
band at $x=9$--10~$\mu$m. (b)~Phase portrait $x$--$p_x$ of the electrons. (c)~The same for the photons.
The light gray band marks the mirror position.}
\label{fig:collision}
\end{figure}

Figure~\ref{fig:collision} shows the collision of the accelerated electrons with the pulse reflected
from the mirror for $n_0=55$, $l=0.06\lambda$ at $t=81$~fs. Some of the electrons located at the front of
the transmitted pulse manage to fly further without taking part in the collision with the reflected
radiation. Therefore, only about 80\% of the electrons with energies above 25~MeV are involved in inverse
Compton scattering. This is seen from a comparison of the electron and photon phase portraits
(Figs.~\ref{fig:collision}b and \ref{fig:collision}c): photons with energies above 100~keV are not
produced by the leading-edge electrons and appear only for the electrons following behind. The
positions of the photon momentum maxima coincide with the maxima of the electron momentum, i.e., the
hardest photons are produced by the most energetic electrons. The second-stage photons are generated
near the mirror surface (Fig.~\ref{fig:collision}a), and their characteristics are shown in
Fig.~\ref{fig:phspec} of the main text.

\section{Angular distribution and solid angle of the radiation}

The energy distribution of the electrons over $p_y/p$ (Fig.~\ref{fig:phspec}c of the main text) has two
maxima at $|p_y/p|\approx0.2$--0.27 ($\vartheta\approx12$--$16^\circ$), caused by the transverse
momentum that the electrons have in the field of the linearly polarized wave (its projection is seen in
the simulation plane). Hard photons with $p_xc>100$~keV follow this two-lobe structure: the maxima of
their distribution (Fig.~\ref{fig:phspec}f of the main text) are located at
$|p_y/p|\approx0.13$--0.17 ($\vartheta\approx8$--$10^\circ$), and 80\% of the energy is emitted within
$|p_y/p|<0.25$. The distribution of softer photons with $p_xc=10$--100~keV is much broader: it is almost
uniform up to $|p_y/p|\approx0.7$ ($\vartheta\approx45^\circ$) and has a dip near the axis. This is a
consequence of the wiggler regime of scattering with $K=a\gg1$: radiation from different parts of the
trajectory is emitted at different angles within $K/\gamma$ in the polarization
plane~\cite{corde_RMP_2013}. Projected onto the simulation plane for $K=a_0$, this angle ranges from
$5^\circ$ at $\gamma=200$ to $27^\circ$ at $\gamma=40$. The hardest photons are emitted by the
electrons with the highest energy, for which this angle is small, while the soft part of the spectrum is
formed by electrons of lower energy.

When estimating the solid angle $\Delta\Omega$ into which the radiation is emitted, one must take into
account that it is not axially symmetric. The electrons oscillate in the polarization plane, which makes
an angle of $45^\circ$ with the simulation plane, and the photons are emitted into a fan elongated along
it, while the distributions in Fig.~\ref{fig:phspec}f of the main text represent its projection onto the
simulation plane. This is also indicated by the sharp cutoff of the distribution of 10--100~keV photons
at $|p_y/p|\approx0.77$, close to $1/\sqrt2\approx0.71$, which would be absent for an axially symmetric
distribution. The thickness of the fan is determined by the spread of the electron momenta across the
polarization plane, which for the recorded trajectories of electrons with energies above 30~MeV is
$\delta p_\perp\approx7m_ec$, compared with a spread of $\approx30m_ec$ along it.

If $\sigma_y$ is the rms value of $p_y/p$ for photons in a given energy interval, then the angular
half-width of the fan along the polarization plane is
$\sigma_\parallel\approx(2\sigma_y^2-\sigma_\perp^2)^{1/2}$, and across it
$\sigma_\perp\approx\delta p_\perp/(\gamma m_ec)$. For a Gaussian distribution, the region containing
80\% of the energy has a solid angle $\Delta\Omega\approx2\pi\ln5\,\sigma_\parallel\sigma_\perp$. With
$\sigma_y=0.21$, $\sigma_\perp\approx0.07$ for photons above 100~keV and $\sigma_y=0.5$,
$\sigma_\perp\approx0.2$ for 10--100~keV photons, this gives $\approx0.2$ and $\approx1.4$~sr,
respectively. To refine the estimate, a test-particle model was used: electrons with momenta taken from
the recorded trajectories collided with a counterpropagating plane wave of amplitude 20--35 with the
polarization of the reflected pulse, and the radiation was calculated from the local synchrotron spectrum
and directed along the electron momentum. The model reproduces the positions of the lobes of the
hard-photon distribution ($|p_y/p|\approx0.2$ versus 0.17 in the kinetic simulation). In the model, the
ratio of the solid angle of the fan to that of an annular cone constructed from the same projection is
0.6--0.7 for soft and 0.4--0.5 for hard photons. Applying this ratio to the distributions obtained in
the kinetic simulation, we find that 80\% of the second-stage radiated energy is contained within a
solid angle of $\approx1.2$~sr (1.1--1.3~sr) for photons with energies of 10--100~keV and
$\approx0.13$~sr (0.11--0.18~sr) for photons above 100~keV. These values are used in the main text to
estimate the brightness. For the emission from all three stages, which is used for comparison with the
experiment of Ref.~\cite{shou_NP_2022}, the same ratio gives $\approx1$~sr above 10~keV and
$\approx0.4$~sr above 100~keV. This estimate is rougher, since the model was constructed for the second
stage.

\end{document}